\documentstyle[aps,prl,preprint,epsf,rotate,amsfonts]{revtex}
\tighten
\begin{document}
\preprint{Applied Physics Reports 97-9}
\title{Interplay of Coulomb blockade and Aharonov-Bohm resonances
in a Luttinger liquid}
\author{Jari M. Kinaret$^1$, Mats Jonson$^1$, Robert I. Shekhter$^1$
\and Sebastian Eggert$^2$}
\address{
$^1$Department of Applied Physics and
$^2$Institute of Theoretical Physics,\\
Chalmers University of Technology and 
G\"oteborg University, S-41296 G\"oteborg, Sweden}
\maketitle
\begin{abstract}
We consider a ring of strongly interacting electrons connected to
two external leads by tunnel junctions. 
By studying the positions of conductance
resonances as a function of gate voltage and magnetic flux the interaction
parameter $g$ can be determined experimentally.
For a finite ring 
the minimum conductance is strongly influenced by device geometry
and electron-electron interactions. In particular, if the tunnel 
junctions are close to
one another the interaction-related orthogonality catastrophe is suppressed
and the valley current is unexpectedly large.
\end{abstract}
\date{\today}

\narrowtext

When the size of an electronic system is reduced, a rich variety of new
``mesoscopic'' phenomena becomes experimentally observable. 
Some of the new phenomena
are essentially classical, owing their existence to the granularity of
the electric charge and the system size -dependence of various energy scales.
The most widely studied example of this type is Coulomb blockade.
A different category of mesoscopic phenomena is entirely quantum mechanical,
and is due to the fact that the phase coherence length at low temperatures
is comparable to the system size, giving rise to a number of interference
effects. A particular example is the existence of a persistent current
in the ground state of a mesoscopic ring. In this Letter we study
the interplay between two mesoscopic phenomena, Coulomb blockade and 
Aharonov-Bohm interference, using an exactly solvable model.

We consider a system consisting of a small ring of interacting electrons
connected to two non-interacting leads by tunnel junctions.
The tunnel junctions are at positions $x_L$ and $x_R$,
respectively. The ring is capacitively coupled to an external gate electrode
and may be pierced by a magnetic flux. We consider a small
AC voltage applied to the right lead and wish to evaluate the current at the 
left junction. A straightforward application of Kubo formula 
\cite{Mahan} yields the current
\begin{displaymath}
\langle I_L(t)\rangle = -i\frac{V(t)}{\hbar\Omega}
\int_{-\infty}^t\!\!\!\!
dt' e^{-i\Omega(t'-t)}{\rm Tr}\left\{\hat\rho_G\left[\hat I_R(t'),
\hat I_L(t)\right]\right\}
\end{displaymath}
where $V(t) = V_0e^{-i\Omega t}$ is the applied voltage and $\hat\rho_G$
is the equilibrium density matrix.
The quantity on the right hand side
is recognized as the retarded 
current-current correlation function. It is most readily evaluated in
imaginary time.

The Hamiltonians for the leads and the 
connection between the leads and the ring
are given by
\begin{eqnarray}
H_L &=& \sum_k\epsilon_{kL}c_{L}^\dagger(k)c_{L}(k)\\
H_{LT} &=& t_Lc_L^\dagger(x_L)\psi(x_L) + t_L^*\psi^\dagger(x_L)c_L(x_L)
\end{eqnarray}
where $c_L$ are operators on the left lead and $\psi$ are operators on the
ring. The Hamiltonians for the right lead are defined analogously. 
Thus, we take the leads to be non-interacting and couple them to the
ring with tunnel junctions at positions $x_L$ and $x_R$.
To calculate the current-current correlation function we define the generating
functional $Z[J_L,J_R] = {\rm Tr}\exp[-\beta H 
- \frac{1}{\hbar}\int_0^{\hbar\beta}
d\tau (J_L(\tau)I_L(\tau) + J_R(\tau)I_R(\tau))]$ and integrate out the 
free fermions in the leads. That yields
\widetext
\begin{equation}
\begin{array}{l}
Z[J_L,J_R] = Z_{L}Z_{R}{\rm Tr}\exp\biggl\{-\beta H_{\rm ring} 
-\frac{1}{\hbar^2}\int_0^{\hbar\beta}d\tau\int_0^{\hbar\beta}d\tau'\\
\biggl[|t_L|^2(1 - \frac{ie}{\hbar}J_L(\tau))(1 + \frac{ie}{\hbar}J_L(\tau'))
\psi^\dagger(\tau,x_L)G_L(\tau-\tau';x_L,x_L)\psi(\tau',x_L)\\
+ |t_R|^2(1 + \frac{ie}{\hbar}J_R(\tau))(1 - \frac{ie}{\hbar}J_R(\tau'))
\psi^\dagger(\tau,x_R)G_R(\tau-\tau';x_R,x_R)\psi(\tau',x_R)
\biggr]\biggr\}
\end{array}
\end{equation}
where $Z_{L/R}$ and $G_{L/R}(\tau;x,x')$ are the partition functions and
free fermion propagators in the leads, and $H_{\rm ring}$ is the
Hamiltonian for an isolated ring.
The imaginary time ordered correlation function 
$\chi(\tau_1-\tau_2) = 
-\langle T_\tau(I_L(\tau_1)I_R(\tau_2))\rangle$
is obtained by differentiating
$Z[J_L,J_R]$ with respect to $J_L$ and $J_R$. 
To simplify the notation we introduce the four-operator expectation 
value $A(\tau_1,\tau,\tau_2,\tau') = A(\tau_1-\tau_2,\tau_1-\tau,\tau_2-\tau')
= \langle T_\tau(\psi^\dagger(\tau,x_L)\psi(\tau_1,x_L)\psi^\dagger(\tau_2,x_R)
\psi(\tau',x_R))\rangle$
and its Fourier transform $A(i\omega_n,i\omega_n',i\omega_n'')$. 
To the lowest nonzero order
in the tunneling matrix elements $\chi$ is given by
\begin{equation}
\begin{array}{ll}
\chi(i\Omega_n) &
\begin{displaystyle}
= \frac{e^2}{\hbar^4}|t_Lt_R|^2\frac{1}{(\hbar\beta)^2}
\sum_{i\omega_n,i\omega_n'}\biggl\{A(i\Omega_n,-i\omega_n,-i\omega_n')
\end{displaystyle}
\\
&
\begin{displaystyle}
{}\times[G_L(i\omega_n-i\Omega_n;x_L,x_L) - G_L(i\omega_n;x_L,x_L)]
[G_R(i\omega_n';x_R,x_R) - G_R(i\omega_n' + i\Omega_n;x_R,x_R)]\biggr\}
\end{displaystyle}
\end{array}
\label{eq:chi}
\end{equation}
\narrowtext
\noindent
If we regard the leads as infinite (rather than semi-infinite), the 
propagators in the leads are easy to evaluate and yield
$G_L(i\omega_n;x_L,x_L) = 
-i\frac{\hbar}{2}D_L(\epsilon_F){\rm sign}(\omega_n)$ where $D_L(\epsilon)$
is the density of states in the left lead.

To evaluate the four-operator product 
$A(i\omega_n,i\omega_n',i\omega_n'')$ we must specify the
Hamiltonian for the ring. We choose to work with the simplest
exactly solvable interacting model, the 
spinless Luttinger model. 
In the bosonized form the Hamiltonian reads\cite{Haldane}
\begin{displaymath}
H_{\rm ring} = \frac{\pi\hbar}{2L}[\frac{v}{g}(\hat N - N_0)^2 + 
gv(\hat J - J_0)^2]
+ \sum_{q\neq 0}\hbar v|q|b_q^\dagger b_q
\end{displaymath}
where $\hat N$ and $\hat J$ are zero modes associated with the total
charge and total current. Since the numbers of clockwise and counterclockwise
moving electrons on the ring must both be integers, the quantum numbers
$N$ and $J$ must satisfy $(-1)^N = (-1)^J$.
The gate voltage and magnetic flux determine the parameters $N_0 = CV_g/e$
and $J_0 = 2\Phi/\Phi_0$ 
which in turn determine the ground state charge and current.
The parameter $g$ is a measure of the interaction strength, and
equals one for non-interacting electrons\cite{Kane}.
For future use we also define the shorthand notation
$\gamma = \frac{1}{2}(g + g^{-1}) - 1$ 
which vanishes in the non-interacting limit.

Due to time-ordering the exact
expression for $A$ is quite complicated although in principle straightforward.
Since we consider only the lowest order in the tunneling
Hamiltonian, our analysis is valid only sufficiently far from the 
conductance resonances. 
Therefore, we can assume
that the ground state is separated from the excited states by an energy
gap $\delta\epsilon$ 
that is larger than $k_BT$. This approximation is basically similar
to the one used by Fazio and co-workers\cite{Fazio} for an interacting 
ring connected to superconducting leads.
We also neglect events with all four imaginary times approximately
equal since their contribution is negligible.
That allows us to evaluate $A$ in an approximate fashion, and gives 
\begin{displaymath}
\begin{array}{l}
A(i\Omega_n,-i\omega_n,-i\omega_n') \approx
\hbar\beta[\delta_{\Omega_n,0}G(i\omega_n,0)G(i\omega_n',0)\\
- \delta_{\Omega_n,\omega_n-\omega_n'}
G(i\omega_n,x_L-x_R)G(i\omega_n-i\Omega_n,x_R-x_L)]
\end{array}
\end{displaymath}
where $G(\tau,x) = -\langle T_\tau(\psi(\tau,x)\psi^\dagger(0,0))\rangle$
is the Green's function for interacting electrons on the ring
and $G(i\omega_n,x)$ is its Fourier transform with respect to the imaginary
time variable $\tau$.
For non-interacting electrons we can apply Wick's theorem and find that this
expression is exact. Substituting this into the expression (\ref{eq:chi})
gives, after proper analytic continuations, the DC conductance
\widetext
\begin{equation}
\renewcommand{\arraystretch}{1.5}
\begin{array}{ll}
\sigma & 
\begin{displaystyle}
\approx \frac{e^2}{h}\frac{|t_Lt_R|^2}{\hbar^2}
D_L(\epsilon_F)D_R(\epsilon_F)
\int_{-\infty}^\infty d\omega 
\left(-\frac{\partial n_F(\omega)}{\partial\omega}\right)
G^{\rm ret}(\omega,x_L-x_R)G^{\rm adv}(\omega,x_R-x_L)
\end{displaystyle}\\
&
\begin{displaystyle}
\approx
\frac{e^2}{h}\frac{|t_Lt_R|^2}{\hbar^2}
D_L(\epsilon_F)D_R(\epsilon_F)
|G^{\rm ret}(\omega = 0,x_L-x_R)|^2
\end{displaystyle}
\end{array}
\label{eq:dccond}
\renewcommand{\arraystretch}{1.0}
\end{equation}
\narrowtext
\noindent
where the last expression is valid if the system is
further than $k_BT$ from a resonance.

This expression can also be understood using a scattering matrix
approach regarding the ring as a (complicated) scatterer for the
free electrons in the leads as discussed in a specific case by
Jagla and Balseiro\cite{Jagla}. From that point of view our basic approximation
is that one scattering event is completed
before another one takes place --- the approximation breaks down near
resonance when the dwell time for extra electrons in the ring is large.

Now we turn to evaluating the retarded Green's function for interacting
electrons in a finite ring at a finite temperature. We use the low-energy
expansion $\psi(x) = \psi_+(x) + \psi_-(x)$ where $\psi_\pm(x)$ are 
clockwise and counterclockwise moving fermions so that 
$G(\tau,x) = G_{++}(\tau,x) + G_{--}(\tau,x)$. 
Following Haldane\cite{Haldane}, 
the correlation functions can be evaluated exactly and
we obtain ($p = \pm 1$)\cite{Mattsson}
\widetext
\renewcommand{\arraystretch}{1.5}
\begin{equation}
\begin{array}{c}
\langle \psi_p(x,\tau)\psi_p^\dagger(0)\rangle =
\frac{i}{2L}e^{-i\frac{p\pi x}{L}}
e^{\frac{\pi}{L}(N_0g^{-1}v + pJ_0gv)\tau}
\\
\frac{
\vartheta_3(ig^{-1}\alpha N_0 - x_N,
e^{-2g^{-1}\alpha})
\vartheta_3(ig\alpha pJ_0 - x_J,
e^{-2g\alpha})
+
\vartheta_2(ig^{-1}\alpha N_0 - x_N,
e^{-2g^{-1}\alpha})
\vartheta_2(ig\alpha pJ_0 - x_J,
e^{-2g\alpha})
}{
\vartheta_3(ig^{-1}\alpha N_0,
e^{-2g^{-1}\alpha})
\vartheta_3(ig\alpha pJ_0,
e^{-2g\alpha})
+
\vartheta_2(ig^{-1}\alpha N_0,
e^{-2g^{-1}\alpha})
\vartheta_2(ig\alpha pJ_0,
e^{-2g\alpha})
}
\\
\frac{1}{\vartheta_1(\frac{\pi(iv\tau - px)}{L},
e^{-\alpha})}
\left|\frac{(a/L)}{2\vartheta_1(\frac{\pi(iv\tau - px)}{L},
e^{-\alpha})}\right|^\gamma
\left|\vartheta_1'(0,e^{-\alpha})\right|^{\gamma + 1}
\end{array}
\label{eq:G}
\end{equation}
\renewcommand{\arraystretch}{1.0}
\narrowtext
\noindent
where $\gamma = \frac{1}{2}(g + g^{-1}) - 1$ and
we introduced the shorthand notation
$\alpha = \frac{\pi v\hbar\beta}{L}$,
$x_N = \frac{\pi}{L}(ig^{-1}v\tau - px)$, 
and $x_J = \frac{\pi}{L}(igv\tau - px)$.
Here $a$ is a short distance cutoff for the interaction, and
is of the order of the lattice spacing.
The Jacobi theta functions\cite{Gradshteyn} 
$\vartheta_3$ and $\vartheta_2$ arise
from the $q=0$ modes with $N$ and $J$ both even or odd, respectively,
and the $\vartheta_1$ arises from the bosons with $q\neq 0$.
The appearance of 
doubly periodic elliptic functions
is natural since the Green's function must be periodic in $x$ and antiperiodic
in $\tau$. 
The Jacobi functions appear also in the partition function of
an isolated ring (essentially the $x$- and $\tau$-independent parts of the
above expression) and therefore in most equilibrium properties
of mesoscopic rings like persistent currents\cite{Loss}.

\begin{figure}
\rotate[r]{
\mbox{
\epsfysize=15cm
\epsfbox{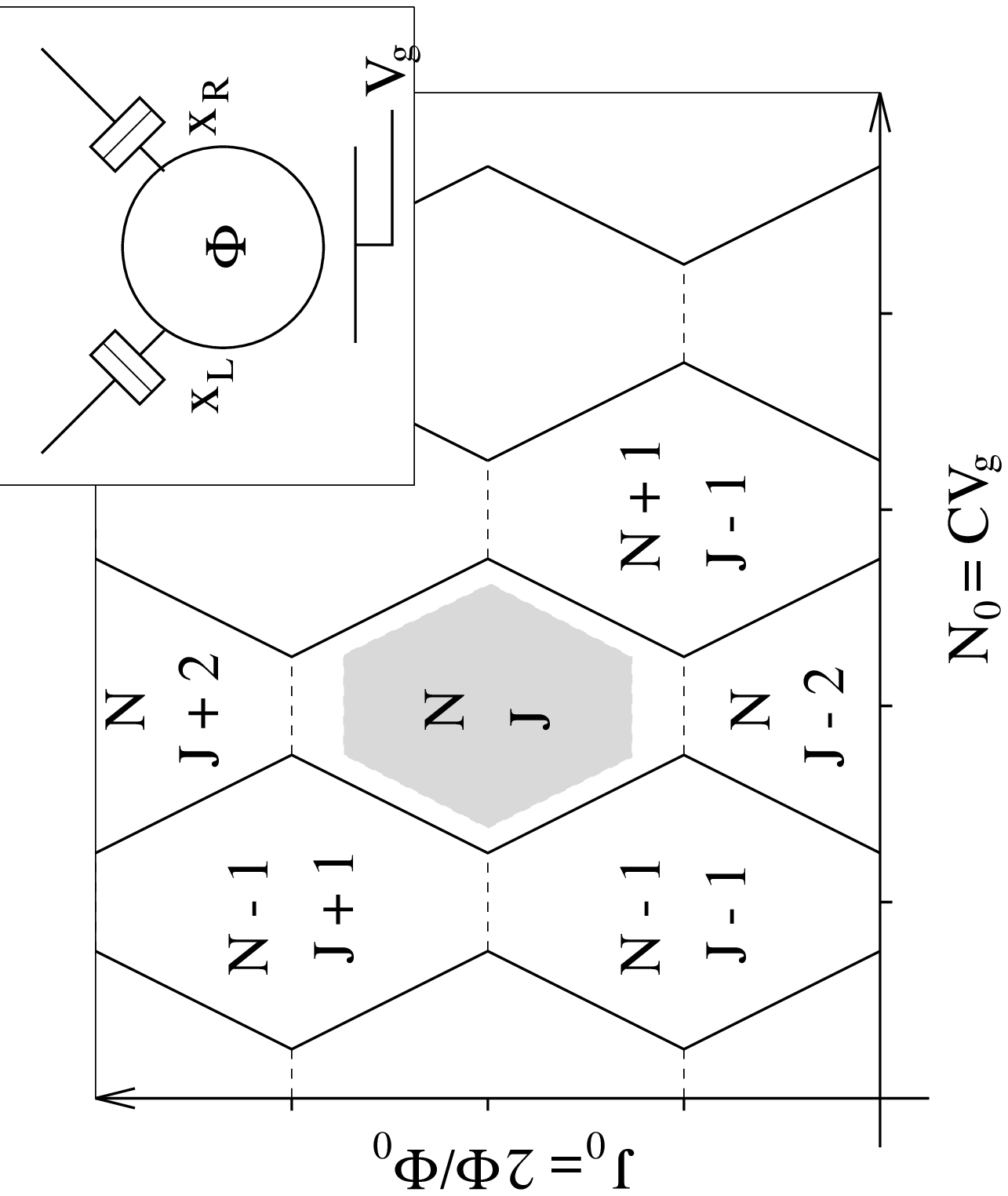}
}
}
\caption{
Positions of conductance resonances in the $(V_g,\Phi)$-plane
for interacting electrons (repulsive
interactions, $g = 1/\protect\sqrt{2}$). 
The labels $N$ and $J$ denote the ground state charge
and current as a function of the external parameters, and the 
shaded area indicates the domain of validity of our analysis.
The line segments with different slopes correspond to 
fluctuations in the numbers of clockwise and counterclockwise moving 
electrons, respectively. Inset: device geometry.
\label{fig:network}
}
\end{figure}

The parameters that are most readily accessible in an experiment are
the gate voltage and the magnetic flux. They enter only the
$q=0$ part of the Hamiltonian
which we can re-write as $H_0 = \frac{1}{2}E_c(\hat N-N_0)^2
+ \frac{\hbar v_F}{2L}(\hat J-J_0)^2$ where $v_F = gv$ is the Fermi velocity
of a non-interacting system with the same density and $E_c = 
\frac{\pi v_F}{g^2L}$ is the charging energy. 
The conductance resonances correspond to values of the gate voltage
and magnetic flux at which the ground state quantum numbers $N$ and $J$
change (degenerate ground state).
Hence, the positions of 
conductance resonances can be determined
from a simple charging energy model with a single-particle Aharonov-Bohm
term --- note, however, that $E_c$ is not simply given by the
geometric capacitance (it is non-zero even for non-interacting electrons).
In the $(V_g,\Phi)$-plane the resonance positions
form a network the shape of which
depends on the interaction parameter $g$.
We suggest therefore that
the interaction parameter can be experimentally measured by studying the
trajectories of conductance maxima as a function of the gate voltage and
magnetic flux.
For non-interacting systems
the resonance positions form a lattice of 
diamond shaped parallelograms, whereas for
repulsive interactions ($g < 1$) there are some values of the gate 
voltage such that a resonance condition is not met for any $\Phi$
as indicated in Figure \ref{fig:network}.
For attractive interactions, there are ranges of $\Phi$ such 
that the total current in the ring, $J$, is independent of $V_g$. In that
case electrons can tunnel into and out of the ring only as pairs of
clockwise- and counterclockwise movers, which is reminiscent of Cooper pair
tunneling through a superconducting grain\cite{Jonson}. 
From now on 
we only consider repulsive interactions.

\begin{figure}
\rotate[r]{
\mbox{
\epsfysize=15cm
\epsfbox{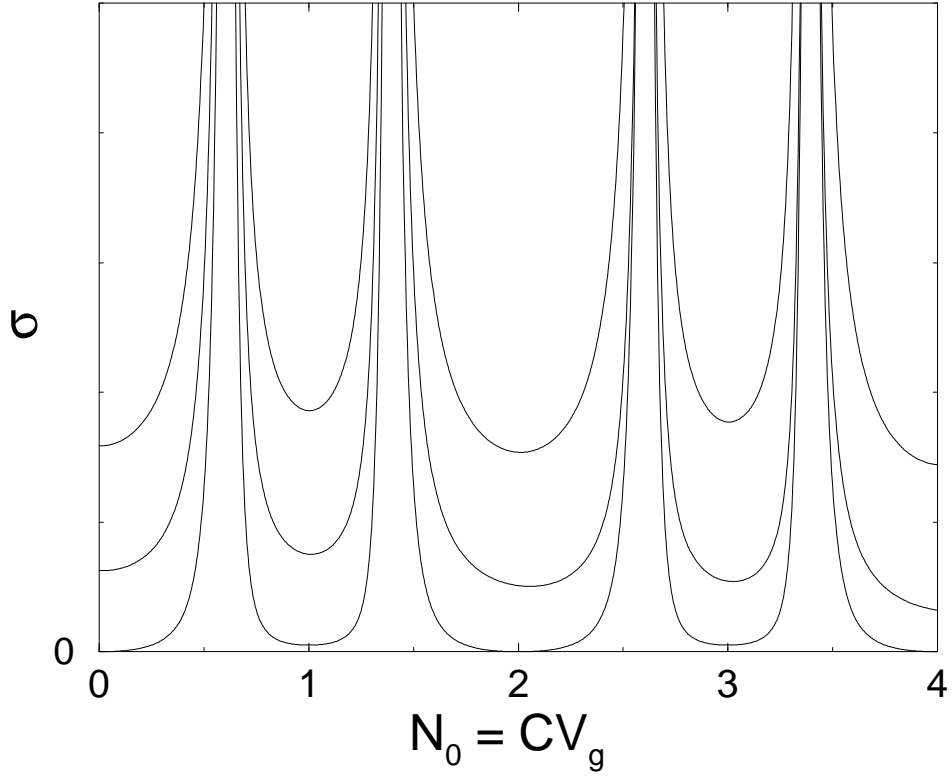}
}
}
\caption{Conductance {\em vs.} $V_g$ for $x_L - x_R = 0.02L$,
$x_L - x_R = 0.05L$ and $x_L - x_R = 0.5L$ (from top to bottom).
The temperature is $T = 0.1\frac{\hbar v}{L}$ and the interaction
parameter is $g=1/2$.
\label{fig:cond}
}
\end{figure}

We use the expressions (\ref{eq:dccond}) and  (\ref{eq:G}) to analyze 
the dependence of the 
conductance on the external parameters. 
The dependences on $V_g$ and $\Phi$ are qualitatively similar
and in Figure \ref{fig:cond} we show the conductance as a function
of $V_g$.
Near a resonance
we obtain the limiting behavior
$|G^{\rm ret}(0,x)|^2 \sim C/(\delta\epsilon)^2$ 
where $\delta\epsilon$ is the energy cost
of changing the number of electrons in the ring by one 
(we still assume $\delta\epsilon > k_BT$).
In a generic case
a  resonance corresponds to a degeneracy for the addition or removal of 
either a clockwise
or counterclockwise moving particle, and the prefactor $C$ is
independent of device geometry. 
However, since we have two control parameters $V_g$ and
$\Phi$, we can use them to bring both clockwise and counterclockwise modes
to resonance simultaneously. These special parameter values
correspond to slope changes in the
trajectories of conductance resonances in Figure \ref{fig:network}. 
Since at these double resonances the Green's function has significant
contributions from wave vectors $k_{F+}$ and $-k_{F-}$, its absolute square
has components with wave vector $k_{F+} - (-k_{F-}) = 2\pi N/L$. Thus, we
find that near a double resonance
the amplitude of the conductance maximum varies periodically
with the separation between the tunnel junctions as
$(1 + \cos[\frac{2N\pi}{L}(x_L - x_R)])$.
Away from a double resonance one channel dominates and 
the amplitude of these oscillations is exponentially small.
Since the wave vector of these oscillations is essentially $2k_F$, they
can be observed only in low-density systems or using local
probes like STM. The interference effects are smeared out by temperature
which leads to a different temperature dependence of the conductance
for different device geometries
near a double resonance: if $N|x_L-x_R| \approx nL$, where $n$ is an
integer, the conductance decreases with temperature due to reduced
interference, whereas for $N|x_L-x_R| \approx 
(n + \frac{1}{2})L$ the conductance increases with $T$.

\begin{figure}
\rotate[r]{
\mbox{
\epsfysize=15cm
\epsfbox{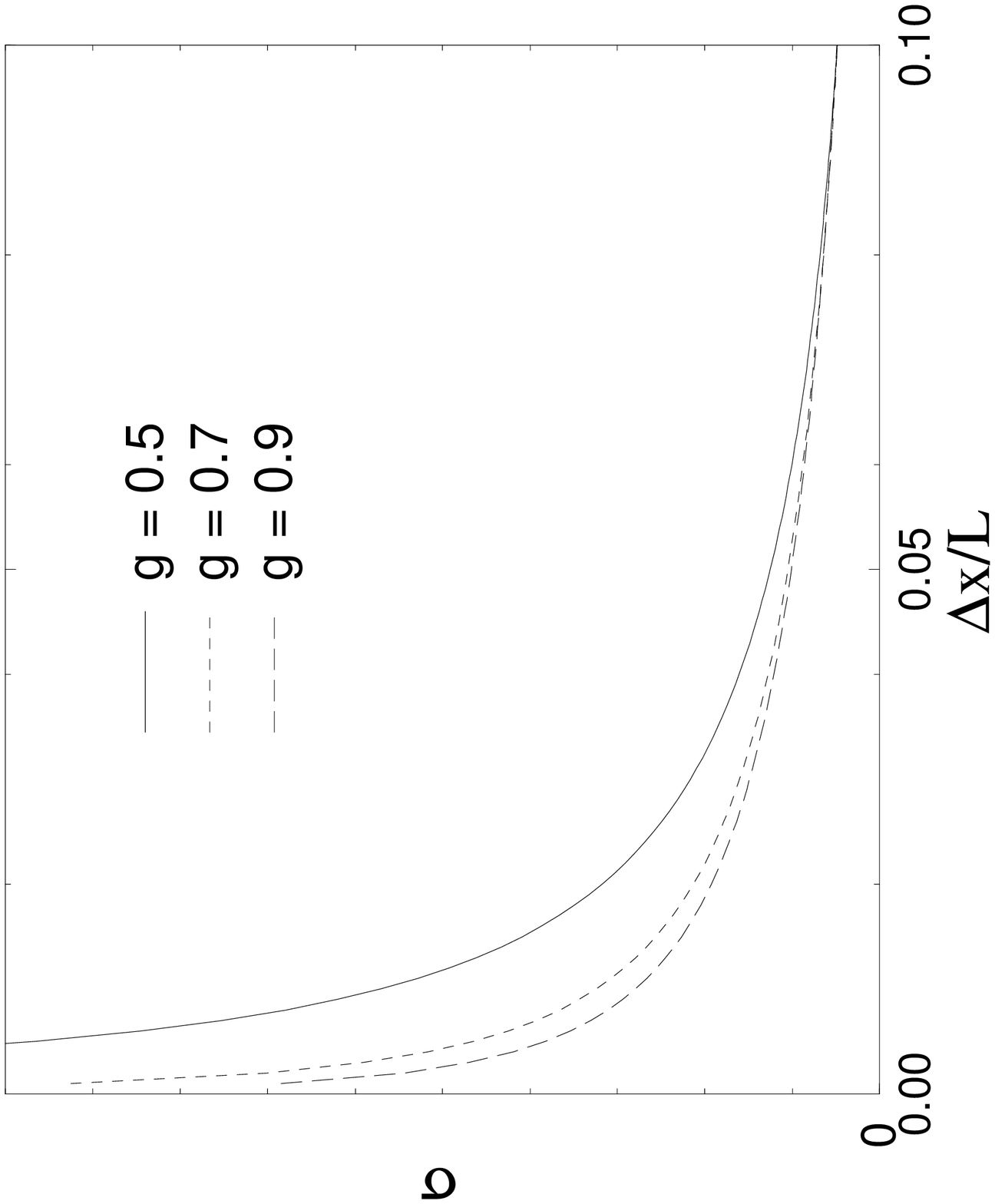}
}
}
\caption{Conductance {\em vs.} $|x_L - x_R|$ near a conductance
minimum for interaction parameters $g=0.9$, $g=0.7$, and $g=0.5$.
The temperature is $T = 0.1\frac{\hbar v}{L}$.
\label{fig:xdep}
}
\end{figure}

From the expression (\ref{eq:G}) we see that the 
conductance
is reduced due to interactions by a factor $(a/L)^{2\gamma}$ which
can be attributed to an orthogonality catastrophe that has previously been
studied in the context of quantum dots in the fractional quantum Hall 
regime \cite{Kinaret}.
The exponent governing the resonance line shape for small $\delta\epsilon$,
$\sigma \sim 1/(\delta\epsilon)^2$,
is independent of the interaction
parameter $g$, which is a consequence of a finite minimum energy
of the bosonic modes. 
The resonant contribution dominates for
$\delta\epsilon \alt \delta\epsilon_c =
\frac{2\pi\hbar v}{L}\left|\sin\left(\frac{\pi(x_L-x_R)}{L}
\right)\right|^{\gamma}$ (up to logarithmic corrections);
for $\delta\epsilon \gg \delta\epsilon_c$ the
valley conductance levels off to a constant value proportional to
$(a/|x_L - x_R|)^{2\gamma}$ as seen in Figure \ref{fig:xdep}.
For large separations $\Delta x = |x_L - x_R|$ the crossover point 
$\delta\epsilon_c$
exceeds half of the resonance spacing and the crossover is not observed.
The two limiting behaviors can be combined to give the approximate
line shape
\widetext
\begin{equation}
\sigma(\delta\epsilon)
\sim \Gamma_L\Gamma_R\left[
\frac{1}{(\delta\epsilon)^2}\left(\sqrt{2}\frac{a}{L}\right)^{2\gamma} +
\left(\frac{L}{2\pi\hbar v}\right)^2
\left|\frac{L}{\sqrt{2}a}\sin\left(\frac{\pi\Delta x}{L}
\right)\right|^{-2\gamma}
\left(\frac{1 - e^{\gamma
\log|\sin(\frac{\pi\Delta x}{L})|}}
{\gamma}
\right)^2\right]
\label{eq:cond}
\end{equation}
\narrowtext
\noindent
where 
$\gamma = \frac{1}{2}(g + g^{-1}) - 1$ and 
$\Gamma_{L/R} = |t_{L/R}|^2D_{L/R}(\epsilon_F)$ are the line widths
for a non-interacting system.
The last factor gives rise to a logarithmic
dependence on $\Delta x$ in the non-interacting limit.

At $T=0$ the smallest $\delta\epsilon$ that we can consider
is determined by when terms that are higher order in the tunneling
Hamiltonian become significant. Due to the orthogonality catastrophe that
happens when the first term in (\ref{eq:cond}) is of order unity.
Therefore, we expect that at $T=0$ the height of the resonance peak
is independent of $g$, and the peak width is given by $\delta\epsilon$
such that $\Gamma_L\Gamma_R\left(\sqrt{2}\frac{a}{L}\right)^{2\gamma}
(\delta\epsilon)^{-2} \approx 1$. 
A simple Breit-Wigner formula $\sigma \sim \frac{\Gamma^2}
{(\delta\epsilon)^2 + \Gamma^2}$ gives a peak-to-valley ratio 
$2(\Gamma/\Delta)^2$ where $\Delta$ is the separation between adjacent
resonances and
$\Gamma$ is the resonance width. In the present case this simple
connection does not hold: the line width at $T=0$ is reduced by
a factor $(a/L)^{\gamma}$ whereas the valley current (for small 
$\Delta x$) is only suppressed by factor $(a/\Delta x)^{2\gamma}$.
The former suppression factor can be identified as the 
life time of a charged excitation of the ring while the second
one is the off-resonance probability of transmission through the ring.
The valley current is therefore anomalously large for small
$\Delta x$ since effectively the system size is replaced by $\Delta x$ 
and the orthogonality catastrophe is less severe.
Since at $T=0$ the peak {\em widths} are reduced by
interactions,
the resonance peak {\em heights} at a finite
temperature are suppressed due to thermal broadening.
Thus, the main effect of interactions is 
to change the peak-to-valley ratio in a way that depends on device geometry
and temperature.

The experimental possibilities for the study of nanostructures like the one
we consider are developing rapidly. New techniques like conducting organic
molecules and carbon compounds are emerging to complement the conventional
semiconductor structures. In particular, it was recently
demonstrated\cite{Liu} that carbon nanotubes exhibit coherent electron
transport and can be used to fabricate
nanoscale ring structures. We believe these devices can be used to
experimentally study the system we have analyzed.

In conclusion, we have considered tunneling through a finite strongly 
interacting system within the framework of an exactly solvable model.
We find that the positions of conductance resonances in the 
$(V_g,\Phi)$-plane can be used to determine the interaction parameter $g$.
We conclude that at $T=0$ the heights of resonance peaks are unaffected
by interactions but due to the narrowness of $T=0$ resonances, the
peak conductance at a finite temperature is reduced by interactions.
The valley current depends on both
interactions and the device geometry. Near a double resonance we find
that the heights of resonance peaks depend on device geometry due to
interference between different current carrying processes.



\begin{references}{}
\bibitem{Mahan}
G.~D.~Mahan, {\em Many Particle Physics}, (Plenum, New York, 1990).
\bibitem{Haldane}
F.~D.~M.~Haldane, J.~Phys.~{\bf C 14}, 2585 (1981).
\bibitem{Kane}
C.~L.~Kane and M.~P.~A.~Fisher, Phys.~Rev.~Lett.~{\bf 68}, 1220 (1992).
\bibitem{Fazio}
R.~Fazio, F.~W.~J.~Hekking, and A.~A.~Odintsov, Phys.~Rev.~Lett.~{\bf 74}, 1843
(1995).
\bibitem{Jagla}
E.~A.~Jagla and C.~A.~Balseiro, Phys.~Rev.~Lett.~{\bf 70}, 639 (1993).
\bibitem{Mattsson}
For details see A.~Mattsson, S.~Eggert, and J.~M.~Kinaret, unpublished
(1997).
\bibitem{Gradshteyn}
I.~S.~Gradshteyn and I.~M.~Ryzhik, {\em Table of Integrals, Series, And 
Products}, (Academic, Orlando, 1980).
\bibitem{Loss}
D.~Loss, Phys.~Rev.~Lett.~{\bf 69}, 343 (1992).
\bibitem{Jonson}
K.~A.~Matveev {\em et al.}, Phys.~Rev.~Lett.~{\bf 70}, 2940 (1993).
\bibitem{Kinaret}
J.~M.~Kinaret {\em et al.}, Phys.~Rev.~B {\bf 45}, 9484 (1992);
{\em ibid}, {\bf 46}, 4681 (1992).
\bibitem{Liu}
J.~Liu {\em et al.}, Nature~{\bf 385}, 780 (1997); S.~J.~Tans {\em et al.},
{\em ibid.} {\bf 386}, 474 (1997).
\end{references}
\end{document}